\documentclass[pra,twocolumn,longbibliography,superscriptaddress]{revtex4-1}

\usepackage{graphicx}
\usepackage{amsmath}
\usepackage{multirow} 
\usepackage{physics} 
\usepackage{braket}
\usepackage{subfigure}   
\usepackage{verbatim}
\usepackage{color}
\usepackage{tikz}

\newcommand{\affa}{State Key Laboratory of Low Dimensional Quantum Physics, Department of Physics, Tsinghua University, Beijing 100084, China}

\newcommand{\affc}{Beijing Academy of Quantum Information Sciences, Beijing 100193, China}
\newcommand{\affd}{Hefei National Laboratory, Hefei 230088, China}
\newcommand{\affe}{Frontier Science Center for Quantum Information, Beijing 100084, China}
\newcommand{\afff}{Department of Physics, The University of Haripur, KP, Pakistan}
\newcommand{\affg}{Centro de Ci\^encias Naturais e Humanas, Universidade Federal do ABC - UFABC, Santo Andr\'e, Brazil}
\newcommand{\affh}{D\'epartement de g\'enie physique, \'Ecole polytechnique de Montr\'eal, Montr\'eal, Canada}
\newcommand{\affIBS}{Center for Trapped Ion Quantum Science, Institute for Basic Science, Daejeon 34126, South Korea}

\begin{document}

\title{Realization of Trapped Ion Dynamics in the Strong-Field Regime and Non-Markovianity}

\author{Kamran Rehan}
\email{Corresponding author: krehan2010@yahoo.com }
\affiliation{\affc}
\affiliation{\affa}
\affiliation{\afff}

\author{Hengchao Tu}
\affiliation{\affa}
\author{Tadeu Tassis}
\affiliation{\affg}
\affiliation{\affh}

\author{Menglin Zou}
\affiliation{\affa}
\author{Zihan Yin}
\affiliation{\affa}
\author{Jing-Ning Zhang}
\affiliation{\affc}

\author{Fernando L. Semi\~ao}
\email{Corresponding author: semiao@gmail.com}
\affiliation{\affg}

\author{Kihwan Kim}
\email{Corresponding author: kimkihwan@ibs.re.kr}
\affiliation{\affc}
\affiliation{\affa}
\affiliation{\affd}
\affiliation{\affe}
\affiliation{\affIBS}

\begin{abstract}
We experimentally investigate trapped-ion dynamics in the strong-driving regime, where the Rabi frequency $\Omega$ is comparable to the vibrational mode frequency $\nu$. In the conventional weak-driving regime ($\Omega \ll \nu$), the dynamics is well described by effective Hamiltonians for the carrier and motional sidebands, associated with detunings $\delta = n\nu$ ($n = 0, \pm 1, \ldots$). In the strong-driving regime ($\Omega \sim \nu$), these interactions can no longer be treated independently. We characterize this physics through the reduced dynamics of the qubit, where non-Markovian behavior emerges as an operational probe of the spin--motion coupling. We observe a structured non-Markovian response, with well-defined maxima that follow the generalized resonance condition $\delta^{2} + \Omega^{2} = \nu^{2}$, reflecting the strong underlying spin--motion hybridization characteristic of the strong-driving regime.
\end{abstract}

\maketitle
\section{Introduction} \label{INT}
Controlling quantum states in atomic systems via laser-induced coherent interactions is central to both quantum optics and quantum information science. In one of the most prominent quantum platforms—laser-cooled trapped ions \cite{leib03,haffner2008quantum}—laser illumination induces coupling between internal (electronic) and external (vibrational) degrees of freedom, typically modeled as a two-level system (qubit) coupled to a harmonic oscillator \cite{leib03,haffner2008quantum,cir95,cmon95,gessner2014local,Nonclassical,superpositions}. In the weak-driving regime, where the laser Rabi frequency $\Omega$ is much smaller than the vibrational frequency $\nu$, the dynamics is well described by effective carrier and sideband Hamiltonians derived within the rotating wave approximation (RWA) \cite{leib03,haffner2008quantum}. This is the basis for a wide range of quantum control protocols.

Decoherence imposes strict limits on the duration of quantum operations, motivating faster coherent control. Increasing the Rabi frequency (driving strength) to values comparable to the trap frequency ($\Omega \sim \nu$) is therefore of interest for both accelerating quantum control and exploring regimes beyond the validity of standard rotating-wave approximations. In this regime, the separation between carrier and motional sidebands no longer holds, and even a laser tuned exactly to the qubit transition can induce qubit--motion coupling~\cite{Cessa}. As a result, the system exhibits strong mixing between internal and motional degrees of freedom, leading to dynamics beyond the conventional weak-driving description. The strong-field regime ($\Omega \sim \nu$) has been theoretically identified in the context of trapped-ion systems as a qualitatively distinct regime of light–matter interaction, with proposals ranging from fast quantum gates \cite{jonathan00,jonathan01} to enhanced motional state reconstruction \cite{tassis23}, and quantum thermal devices \cite{tassis25, Ribeiro}. Despite this rich theoretical interest, accessing this regime experimentally remains challenging due to stringent control requirements. The results reported here provide direct experimental access to this regime.

In the strong-field regime introduced above, the strong hybridization between internal and motional degrees of freedom leads to dynamics that cannot be described within the standard carrier–sideband framework. When viewed at the level of the reduced qubit dynamics, this coherent spin–motion coupling gives rise to non-Markovian (NM) behavior—widely studied in open quantum systems as arising from memory effects and information backflow \cite{rivas10, breuer09, rajagopal10, bylicka14, pineda16, fanchini14, alipour12, dhar15, wolf08, utagi20b, Moreira,Cardenas}. Here, NM emerges intrinsically from the underlying unitary dynamics through the coherent exchange of information between internal and motional degrees of freedom, and thus provides an operational probe of the hybridized regime. 

From this perspective, the qubit can be described as an effective open system, where the motional degrees of freedom act as a coherent environment, complemented by residual Markovian dephasing. In the absence of spin–motion coupling, the qubit undergoes simple exponential dephasing at an effective rate $\gamma$. By contrast, when $\Omega \sim \nu$, coherent spin–motion interactions induce memory effects and lead to pronounced non-Markovian dynamics. Experimentally, we access this regime and characterize the qubit dynamics via spin-state tomography, revealing clear signatures of strong spin–motion coupling through information backflow.

The remainder of this paper is organized as follows. In Sec.~\ref{TS}, we introduce the physical system and describe the experimental setup. Section~\ref{NMS} presents the measure used to quantify non-Markovianity. The experimental results and their analysis are discussed in Sec.~\ref{RST}. Finally, Sec.~\ref{C} summarizes our main findings and presents our conclusions. Details of the numerical simulations and experimental protocols, including statistical analysis, are provided in Appendices~\ref{sim} and~\ref{prot}, respectively.

\section{The system} \label{TS}
We first recall the standard Hamiltonian describing a two-level trapped ion interacting with a classical laser field. The ion is confined in a harmonic trap with frequency $\nu$, and its center-of-mass motion is described by the bosonic operators $a^\dagger$ and $a$. For electric dipole (or quadrupole) transitions, the system is governed by \cite{leib03}
\begin{equation}\label{H}
H=\frac{\hbar \omega_0}{2} \sigma_z + \hbar \nu a^\dagger a + \frac{\hbar \Omega}{2} \left( \sigma_+ e^{i \eta (a + a^\dagger)-i\omega_L t} + \text{h.c.} \right),
\end{equation}
where $\omega_0$ is the internal transition frequency, $\omega_L$ is the laser frequency, $\Omega$ is the Rabi frequency, and $\eta$ is the Lamb–Dicke parameter \cite{leib03}. Here, $\sigma_z=|e\rangle\langle e|-|g\rangle\langle g|$, $\sigma_+=|e\rangle\langle g|$, and $\sigma_-=(\sigma_+)^\dag$. By applying the unitary transformation $U = e^{i \omega_L t \sigma_z / 2}$, the Hamiltonian becomes time-independent,
\begin{equation}\label{Hi}
H = \frac{\hbar \delta}{2} \sigma_z + \hbar \nu a^\dagger a + \frac{\hbar \Omega}{2} \left( \sigma_+ e^{i \eta (a + a^\dagger)} + \text{h.c.} \right),
\end{equation}
with the detuning $\delta = \omega_0 - \omega_L$.

For our experimental system, we consider a single $^{171}$Yb$^+$ in an elliptical Paul trap, with the qubit encoded in the hyperfine manifold of the electronic ground state. The computational basis is defined as
\begin{eqnarray}
\ket{g} &\equiv& \ket{^2S_{1/2}, F = 0, m_F = 0}, \nonumber \\
\ket{e} &\equiv& \ket{^2S_{1/2}, F = 1, m_F = 0}.
\end{eqnarray}

The interaction between these internal states and the motional degrees of freedom is implemented through stimulated Raman transitions driven by two laser beams. In this case, Eq.~(\ref{Hi}) remains valid, with $\delta$ defined relative to the Raman beat-note frequency and $\Omega$ corresponding to the effective Raman Rabi frequency determined by the Raman beam parameters \cite{an2015experimental,um2016phonon,zhang2018noon,shen2018quantum,lv2018quantum,chen2021quantum,zhang2024single,leib03}.

In contrast to the single-mode description introduced above, the anisotropic confinement of our trap gives rise to two radial motional modes with frequencies $\nu_1$ and $\nu_2$, corresponding to vibrations along the $x$ and $y$ directions. The resulting Hamiltonian is
\begin{eqnarray}\label{realH}
H &=& \frac{\hbar \delta}{2} \sigma_z + \sum_{i=1,2} \hbar \nu_i a_i^\dagger a_i \nonumber\\
&&+ \frac{\hbar \Omega}{2} \sum_{i=1,2} \left[ \sigma_+ e^{i \eta_i (a_i + a_i^\dagger)} + \mathrm{h.c.} \right],
\end{eqnarray}
where $\nu_i$ and $\eta_i$ denote, respectively, the frequencies and Lamb–Dicke parameters of the two radial motional modes ($i=1,2$). 

To account for decoherence effects, we describe the dynamics using the master equation
\begin{equation}\label{eq:master_equation}
\frac{d\rho}{dt} = -\frac{i}{\hbar}[H, \rho] + \gamma \left( \sigma_z \rho \sigma_z - \rho \right),
\end{equation}
where $\gamma$ is an effective pure dephasing parameter that phenomenologically captures the effects of residual noise in the experimental system.

\begin{figure}[t]
\centering
\begin{tikzpicture}
\node {\includegraphics[width=0.4\textwidth]{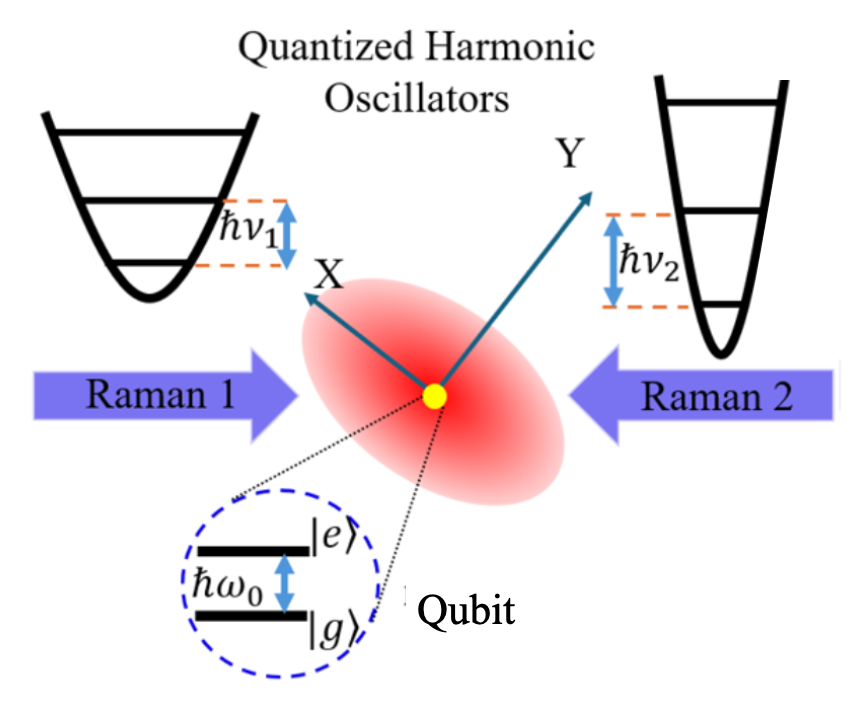}};
\end{tikzpicture}
\caption{Schematic of the trapped-ion system realizing the strong-driving regime.
An elliptical Paul trap produces two radial motional modes with frequencies $\nu_1$ and $\nu_2$, corresponding to vibrations along the $x$ and $y$ directions.
Two internal electronic states define a qubit, which is coupled to the motional modes via a pair of counter-propagating 355~nm Raman beams.
These fields induce an effective spin–motion interaction with tunable strength $\Omega$, enabling access to the strong-field regime $\Omega \sim \{\nu_1,\nu_2\}$.}
\label{fig:Spin-Vibration}
\end{figure}

\section{Non-Markovianity Measure} \label{NMS}
In this work, we analyze the reduced qubit dynamics for different driving and detuning parameters $(\Omega,\delta)$. Details on the reconstruction of the qubit density matrix are provided in Appendix \ref{prot}.

The Breuer–Laine–Piilo measure of non-Markovianity \cite{breuer16}, used here to characterize the dynamics of the qubit, is based on the trace distance between two density operators $\rho_1$ and $\rho_2$, defined as
\begin{equation}
D(\rho_1(t),\rho_2(t))=\tfrac{1}{2}\mathrm{Tr}\,|\rho_1(t)-\rho_2(t)|,
\end{equation}
where $|A|=\sqrt{A^\dagger A}$. This quantity quantifies the distinguishability between quantum states; therefore, any increase in $D$ is interpreted as an information backflow from the environment to the system \cite{wittemer18,breuer16}. Equivalently, non-Markovian behavior is associated with time intervals in which the rate of change
\begin{equation}\label{sigma}
\sigma(t) = \frac{d}{dt}D(\rho_1(t),\rho_2(t)),
\end{equation}
is positive.

Based on this idea, the measure used in our work is defined as \cite{breuer16}
\begin{equation}\label{NM}
\mathrm{NM} = \max_{\rho_1,\rho_2} \int_{\sigma>0} dt\, \sigma(t),
\end{equation}
where the maximization is performed over all pairs of orthogonal pure initial qubit states $\rho_{1,2}(0)$, and the integral is restricted to the time intervals for which $\sigma(t)>0$.
In our experimental implementation, practical constraints require the use of fixed pairs of orthogonal initial states. In particular, we choose eigenstates of $\sigma_x$, which provide an experimentally accessible characterization of non-Markovian features in the qubit dynamics, as discussed in Appendix \ref{sim}. 

\section{Results} \label{RST}
Qubit initialization and readout are performed using a 369.5~nm laser, which enables Doppler cooling, optical pumping, and state detection. Coherent control is implemented via a pair of counter-propagating 355~nm Raman beams, which implement the spin–motion coupling. After Doppler cooling, the motional modes are prepared near their ground state, with mean occupations $\bar{n}_i \sim 0.05$. Additional details on the experimental setup and mode structure are provided in Ref.~\cite{cyangthesis_supp}. 

In our experiments, we observed that the dephasing rate $\gamma$ in Eq.~(\ref{eq:master_equation}) is particularly sensitive to trap characteristics, such as the frequencies $\nu_1$ and $\nu_2$, as well as to the initial qubit state. This behavior is characteristic of an effective dephasing arising from a combination of effects such as power-induced heating, photon scattering, light-shift fluctuations caused by laser-power noise, and motional decoherence~\cite{tu2025polarization}. A careful experimental characterization revealed that, for our current setup, the dephasing rates are $\gamma_+ = 0.0049~\text{MHz}$ for $\rho_{1}(0)=\ket{+}\bra{+}$ and $\gamma_- = 0.0008~\text{MHz}$ for $\rho_{2}(0)=\ket{-}\bra{-}$.

\begin{figure}[h]
    \centering
    \includegraphics[width=0.48\textwidth]{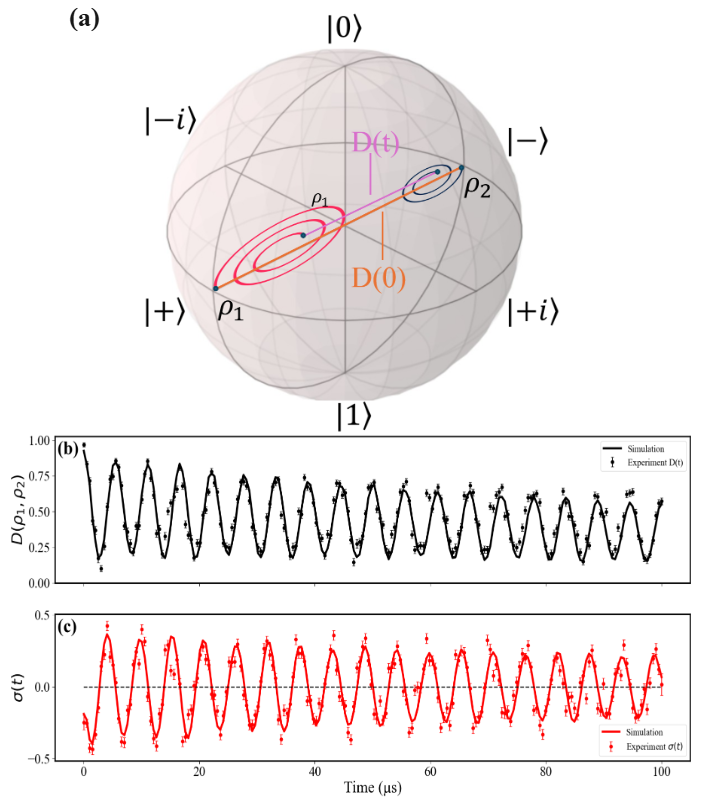}
\caption{Illustration and experimental characterization of the qubit dynamics in the strong-driving regime.
(a) Schematic Bloch-sphere evolution of two initially orthogonal states, $\ket{+}$ and $\ket{-}$, showing the loss and partial recovery of distinguishability induced by spin--motion coupling. 
(b) Time evolution of the trace distance $D(\rho_1,\rho_2)$ between the two states. The overall decay reflects decoherence, while clear nonmonotonic features indicate information backflow from the motional degrees of freedom to the spin. Experimental data (points) are compared with numerical simulations (solid lines). 
(c) Time derivative $\sigma(t)$ of the trace distance. Positive regions (above the dashed line) identify intervals of increasing distinguishability and thus provide the elementary contributions to the non-Markovianity measure.}
\label{fig:trace_distance_and_gradient}
\end{figure}

A schematic representation of the reduced qubit dynamics [Fig.~\ref{fig:trace_distance_and_gradient}(a)] illustrates the coherent exchange between internal and motional degrees of freedom, which gives rise to non-Markovian behavior. This is captured by the trace distance $D(\rho_1,\rho_2)$, whose non-monotonic evolution is shown in Fig.~\ref{fig:trace_distance_and_gradient}(b), providing direct evidence of information backflow. The corresponding time derivative $\sigma(t)$ [Fig.~\ref{fig:trace_distance_and_gradient}(c)] identifies the specific time intervals where this backflow occurs.

The theoretical curves shown in Figs.~\ref{fig:trace_distance_and_gradient}(b) and \ref{fig:trace_distance_and_gradient}(c) are obtained from the master equation [Eq.~(\ref{eq:master_equation})] with Hamiltonian~(\ref{realH}), using the experimentally determined dephasing rates together with $\Omega/(2\pi)=2.245~\text{MHz}$ and $\nu_1(\nu_2)/(2\pi)=2.32(3.16)~\text{MHz}$. They are in good agreement with the experimental data. In Appendix~\ref{sim}, we discuss the robustness of the non-Markovianity measure to variations in the initial qubit and motional states.

\begin{figure}[htbp]
    \centering
    \includegraphics[width=0.46\textwidth]{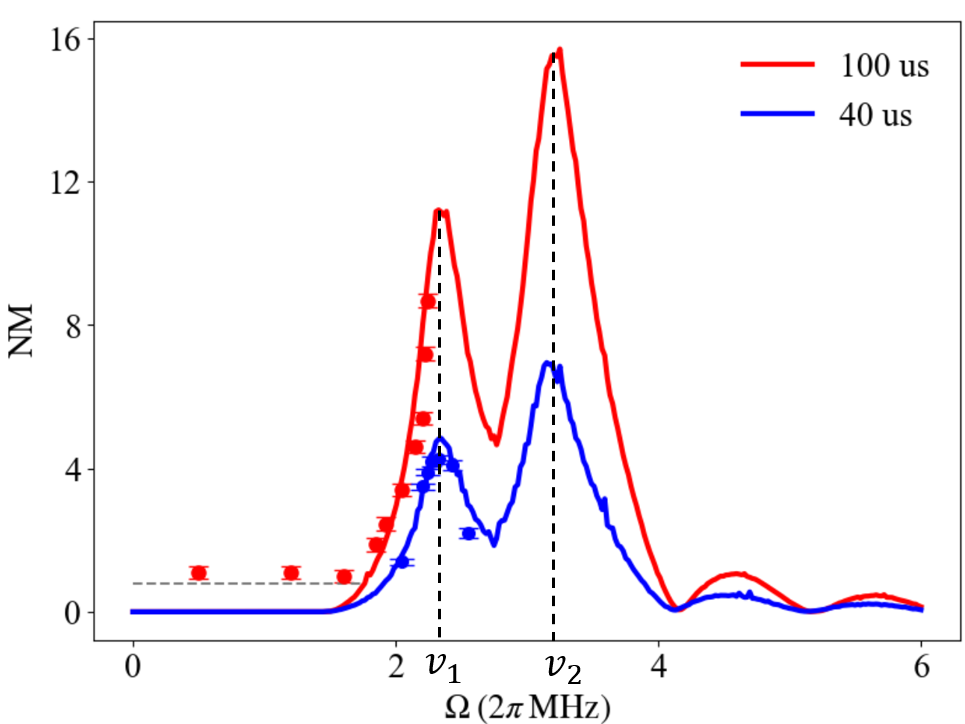}
\caption{Non-Markovianity at resonance ($\delta=0$) as a function of the driving strength. Clear maxima occur near $\Omega=\nu_i$, signaling the onset of resonant spin--motion coupling. The red (upper) and blue (lower) datasets correspond to evolution times of $100~\mu s$ and $40~\mu s$, respectively. The shorter evolution time was required to extend the experimentally accessible range of driving strengths because of experimental limitations (see text). Experimental data (points) are compared with numerical simulations (solid lines).}
\label{fig:New-NM_vs_omega}
\end{figure}

Having identified signatures of non-Markovian behavior, we now investigate its dependence on the driving strength. We focus first on the resonant case $\delta=0$. In the Lamb--Dicke and weak-driving regimes, the dynamics is dominated by a pure carrier interaction \cite{leib03}, where the motional state remains decoupled from the internal dynamics. To systematically explore the dynamics beyond the carrier regime, we evaluate the non-Markovianity measure NM as a function of the Rabi frequency $\Omega$. Figure~\ref{fig:New-NM_vs_omega} shows that NM exhibits a pronounced non-monotonic dependence on $\Omega$: it increases as $\Omega$ approaches the motional frequencies $\nu_i$, signaling the onset of resonant spin--motion coupling, and decreases for larger $\Omega$ as the system moves away from this resonant condition, leading to reduced mixing between internal and motional degrees of freedom. The maxima in NM occur near $\Omega=\nu_i$, where the driven ion realizes an effective Jaynes--Cummings-type interaction with strongly hybridized eigenstates~\cite{Cessa}. In this regime, enhanced information backflow arising from spin--motion coupling provides a physical signature of the breakdown of the carrier approximation and the emergence of dressed-state dynamics.

The measurements at the largest driving strengths were performed using a shorter evolution time of $40~\mu$s instead of $100~\mu$s. This reduction was necessary because, at large $\Omega$, the probability of second ionization and the decay of the Rabi oscillation contrast limit the experimentally accessible evolution time. Although the shorter integration window slightly reduces the accumulated non-Markovianity, it enables reliable measurements over an extended range of driving strengths.

\begin{figure}[h]
    \centering
    \includegraphics[width=0.47\textwidth]{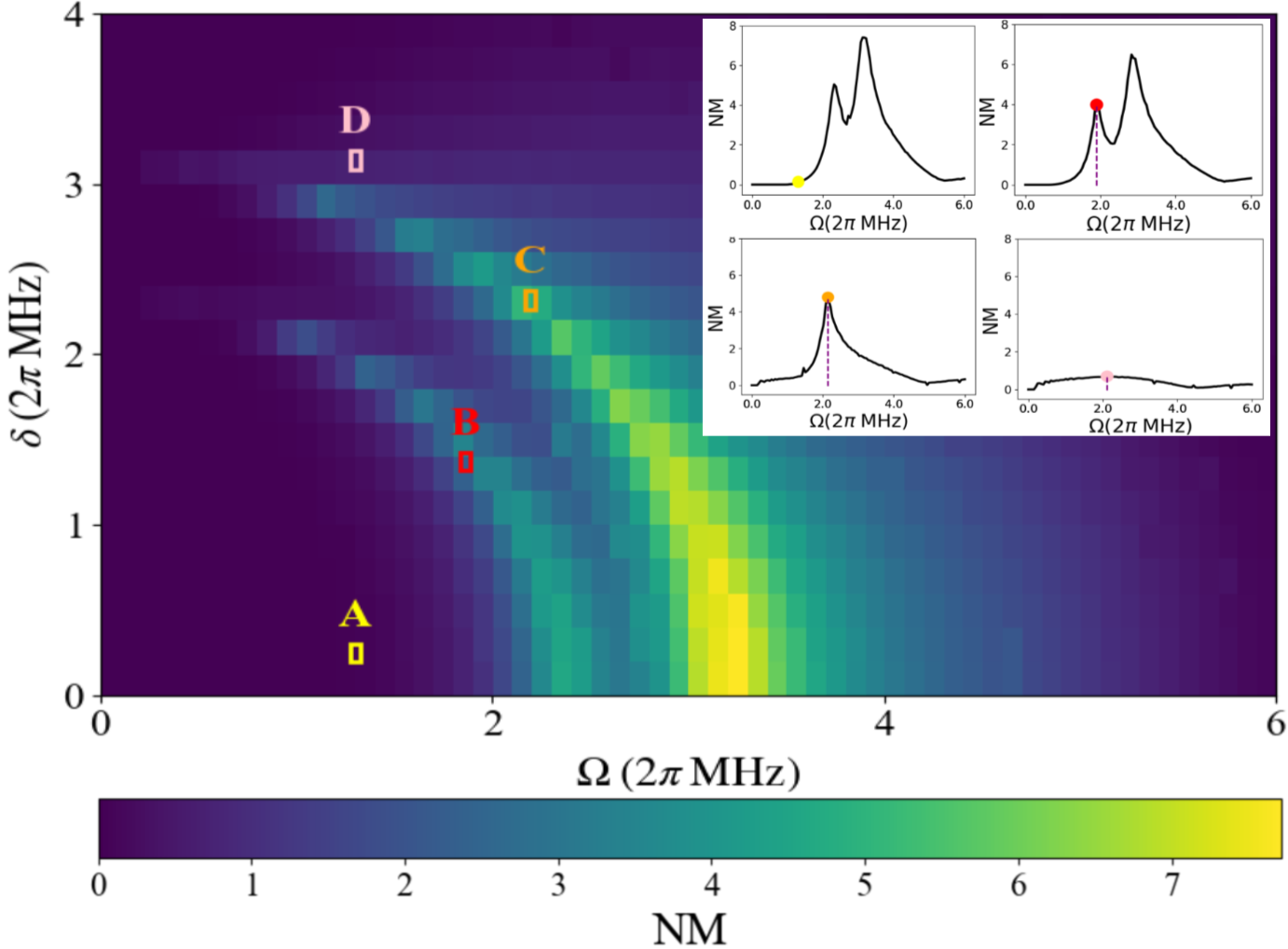}    
\caption{Non-Markovianity in the $(\delta,\Omega)$ plane reveals a structured pattern governed by the generalized resonance condition $\delta^2 + \Omega^2 = \nu_i^2$ $(i=1,2)$. 
Four representative points (A--D), corresponding to detunings $\delta = 0.25$, 1.37, 2.32, and 3.16~MHz, are indicated. 
Inset: NM as a function of $\Omega$ for the selected detunings, showing good agreement between experimental data and theoretical predictions. 
Measurements were performed for a total evolution time of $40~\mu\text{s}$ with 200 time steps.}
\label{fig:New2-circular_pattern}
\end{figure}

The non-monotonic dependence of NM on $\Omega$ observed in Fig.~\ref{fig:New-NM_vs_omega} suggests an underlying resonance structure in the strong-driving regime. Theory predicts that, in the conventional weak-driving regime for a single motional mode, the resonances follow the carrier and sideband conditions $\delta=n\nu$ ($n=0,\pm1,\ldots$) \cite{leib03}. In contrast, in the strong-driving regime, the resonance condition is generalized to $\delta^2+\Omega^2=\nu^2$ \cite{tassis23}. To determine whether the NM also reflects this generalized resonance structure, we extend the analysis to finite detuning $\delta$. Figure~\ref{fig:New2-circular_pattern} shows that, for our system with two radial motional modes, the NM maxima are found along the curves $\delta^2+\Omega^2=\nu_i^2$ $(i=1,2)$, forming circular features in the $(\delta,\Omega)$ plane. These resonant curves define loci of enhanced spin--motion coupling, which are directly reflected in the non-Markovianity of the reduced qubit dynamics. In Fig.~\ref{fig:New2-circular_pattern} we highlight four representative points (A--D), corresponding to $\delta=0.25$, 1.37, 2.32, and 3.16~MHz. The corresponding cuts of NM as a function of $\Omega$ (inset) are in good agreement with theory, demonstrating that NM provides a sensitive probe of the generalized resonance structure.

\section{Conclusions} \label{C}
We have experimentally accessed the strong-driving regime of laser--ion interactions, where the Rabi frequency becomes comparable to the motional frequency ($\Omega\sim\nu$), beyond the validity of the conventional carrier and sideband picture. In this regime, the dynamics is governed by strong spin--motion hybridization. We show that this coupling is encoded in the reduced qubit dynamics through pronounced non-Markovian behavior. The observed information backflow provides an operational signature of spin--motion mixing, exhibiting a non-monotonic dependence on the Rabi frequency and well-defined maxima along the generalized resonance condition $\delta^2+\Omega^2=\nu^2$. Our results demonstrate that this condition, previously identified theoretically, can be experimentally resolved through the reduced qubit dynamics.

These findings establish non-Markovianity as a sensitive probe of the strong-driving regime and its underlying spin--motion hybridization. More broadly, they reveal the crossover from the conventional weak-driving sideband picture to the generalized resonance structure of the strong-driving Hamiltonian, showing that the latter can be directly mapped through the non-Markovian dynamics of the qubit. Our work thus opens a route to diagnosing and harnessing system--environment correlations in quantum platforms.

\section*{acknowledgments}
This work was supported by the Innovation Program for Quantum Science and Technology under Grants No. 2021ZD0301602 and the National Natural Science Foundation of China under Grants No. 62335013, 92576204, and 92565306. T.T. acknowledges partial support from Coordena\c{c}\~ao de Aperfei\c{c}oamento de Pessoal de N\'ivel Superior (CAPES, Finance Code 001). F. L. S. acknowledges partial support from  Funda\c c\~ao de Amparo a Pesquisa do Estado de S\~ao Paulo (FAPESP) Process No. 2021/14135-1 and CNPq (Grant No. 313068/2023-2).
KK acknowledges the support by the IBS Grant No. IBS-R041-D1.

K.R. and H.T. contributed equally to this work.


\appendix

\makeatletter
\counterwithin{figure}{section}
\counterwithin{table}{section}
\makeatother

\section{Experimental Protocols and Statistics}\label{prot}

\subsection{State Reconstruction}

The system is initialized in a product state $\rho_S(0)\otimes\rho_E(0)$, with the qubit prepared in a pure state and the motional modes cooled close to the ground state. The subsequent dynamics is driven by shining two laser beams onto the ion to induce stimulated Raman transitions, characterized by a tunable effective Rabi frequency $\Omega$, over a variable interaction time $t$.

\begin{figure}[h]
\centering
\includegraphics[width=0.95\columnwidth]{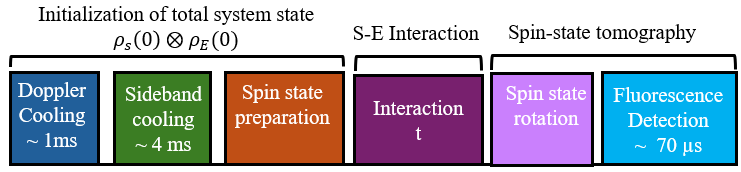}
\caption{Experimental sequence for state reconstruction.}
\label{fig:seq}
\end{figure}

After each time step, projective measurements are performed on the qubit observables to obtain $\langle \sigma_{x,y,z}(t)\rangle$. Typically, $\sim 600$ experimental repetitions per data point are used to ensure statistical convergence. These measurements are used for state tomography, as summarized in Fig.~\ref{fig:seq}.

\subsection{Non-Markovianity Dependence on Control Parameters}
\begin{figure}[h]
    \centering
    \includegraphics[width=0.47\textwidth]{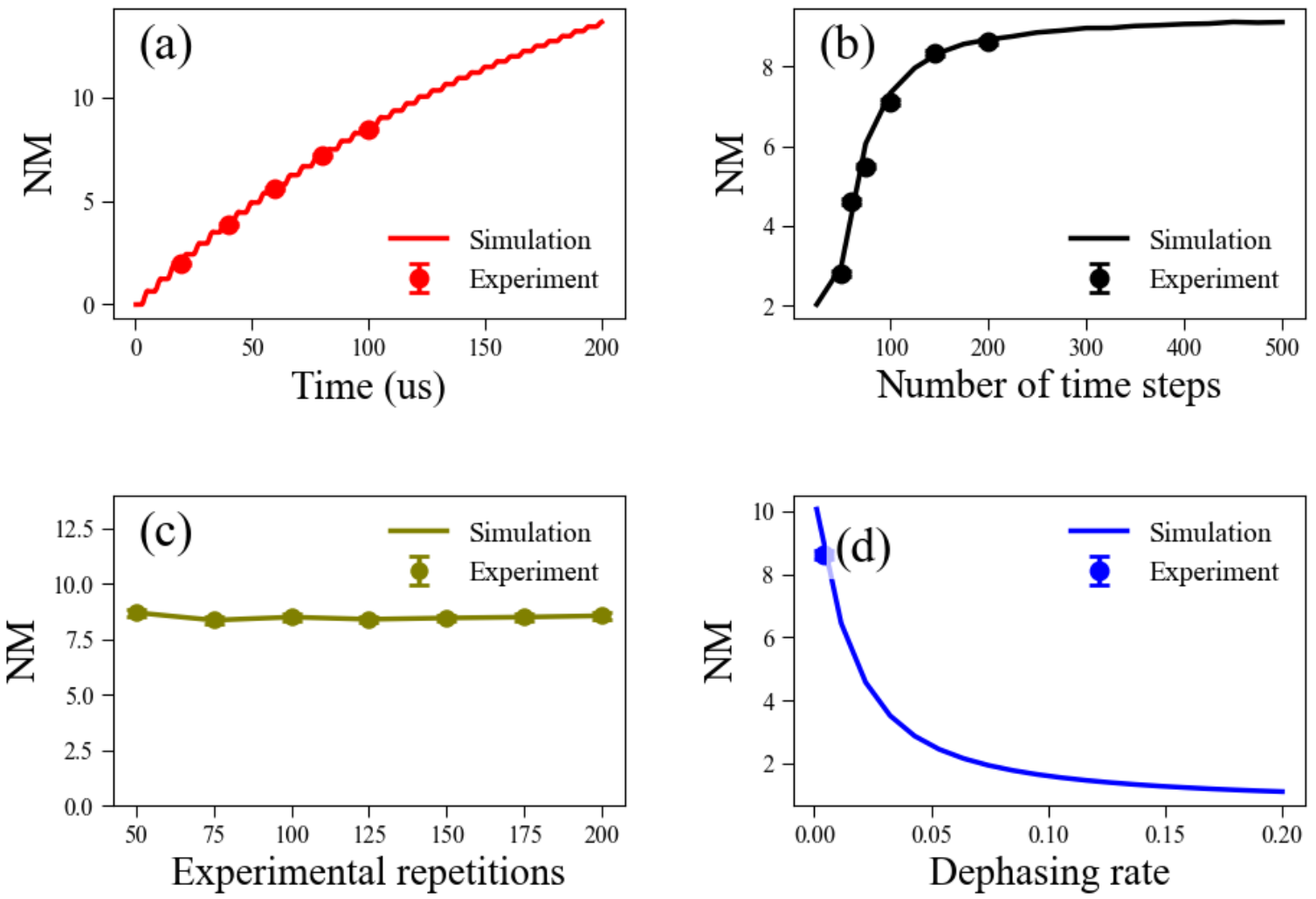}
\caption{Characterization of the experimental protocol, establishing the robustness and convergence of the non-Markovianity estimation. 
(a) NM as a function of the total evolution time. 
(b) Dependence of NM on the number of time steps for a fixed total time of $100~\mu\text{s}$. 
(c) Convergence of NM with the number of experimental repetitions. 
(d) Simulated NM as a function of the dephasing rate of the plus state, with the minus-state dephasing fixed at $0.0008~\text{MHz}$. The marker indicates the experimental value.}
\label{fig:Exp_parameters}
\end{figure}

The dependence of NM [Eq.~(\ref{NM})] on key experimental parameters is summarized in Fig.~\ref{fig:Exp_parameters}. As expected, NM increases over time with a slightly decreasing growth rate, consistent with the damping of $\sigma(t)$ caused by decoherence~[Fig.~\ref{fig:Exp_parameters}(a)]. Consequently, a time window of $100~\mu\text{s}$ is used for all measurements. The time-step analysis [Fig.~\ref{fig:Exp_parameters}(b)] reveals convergence beyond 200 points ($0.5~\mu\text{s}$ resolution), a value adopted throughout this work. Figure~\ref{fig:Exp_parameters}(c) confirms the stability of the extracted NM beyond 50 experimental cycles; nevertheless, more than 500 repetitions were used to ensure statistical robustness. Finally, the dependence on the dephasing rates $\gamma_+$ and $\gamma_-$ [Fig.~\ref{fig:Exp_parameters}(d)] demonstrates the expected behavior under the addition of Markovian noise.

\subsection{Statistical Error Analysis}
For each qubit observable $\langle \sigma_l \rangle \in [-1, 1]$, the statistical uncertainty arising from quantum projection noise (QPN) is estimated assuming binomial statistics,
\begin{equation}
\delta \langle \sigma_l \rangle = 2 \sqrt{\frac{p(1 - p)}{N}},
\label{eq:sigma_error}
\end{equation}
where $p = (\langle \sigma_l \rangle + 1)/2$ and $N$ denotes the number of experimental repetitions~\cite{wittemer18}.
The trace distance between two single-qubit states $\rho_1$ and $\rho_2$ is given by $D = \frac{1}{2} \mathrm{Tr}| \rho_1 - \rho_2 |$, with its uncertainty $\delta D$ obtained via standard error propagation. The time derivative $\sigma(t) = dD/dt$ is evaluated numerically using a symmetric finite-difference scheme, with associated uncertainty
\begin{equation}
\delta \sigma(t) = \frac{1}{\Delta t} \sqrt{ \delta D(t+\Delta t)^2 + \delta D(t-\Delta t)^2 }.
\label{eq:delta_sigma}
\end{equation}
Finally, the non-Markovianity measure $NM= \sum_{\sigma(t) > 0} \sigma(t)\, \Delta t$ carries an uncertainty given by
\begin{equation}
\delta NM = \sqrt{ \sum_{\sigma(t) > 0} \left( \delta \sigma(t)\, \Delta t \right)^2 }.
\label{eq:delta_N}
\end{equation}


\section{Numerical Simulations}\label{sim}

\subsection{Robustness}

In this section, we assess the robustness of the non-Markovianity measure with respect to the choice of initial states. For simplicity we focus on the case of just one motional mode given by the Hamiltonian in Eq.~(\ref{Hi}). While the main text focuses on eigenstates of $\sigma_x$, we compare these results with those obtained for eigenstates of $\sigma_y$ and $\sigma_z$. We find consistent qualitative behavior across different initial spin configurations, with only minor quantitative variations. We also examine the dependence on the initial motional state and observe that the main features are preserved. 

\begin{figure}[h]
\centering
\includegraphics[width=\linewidth]{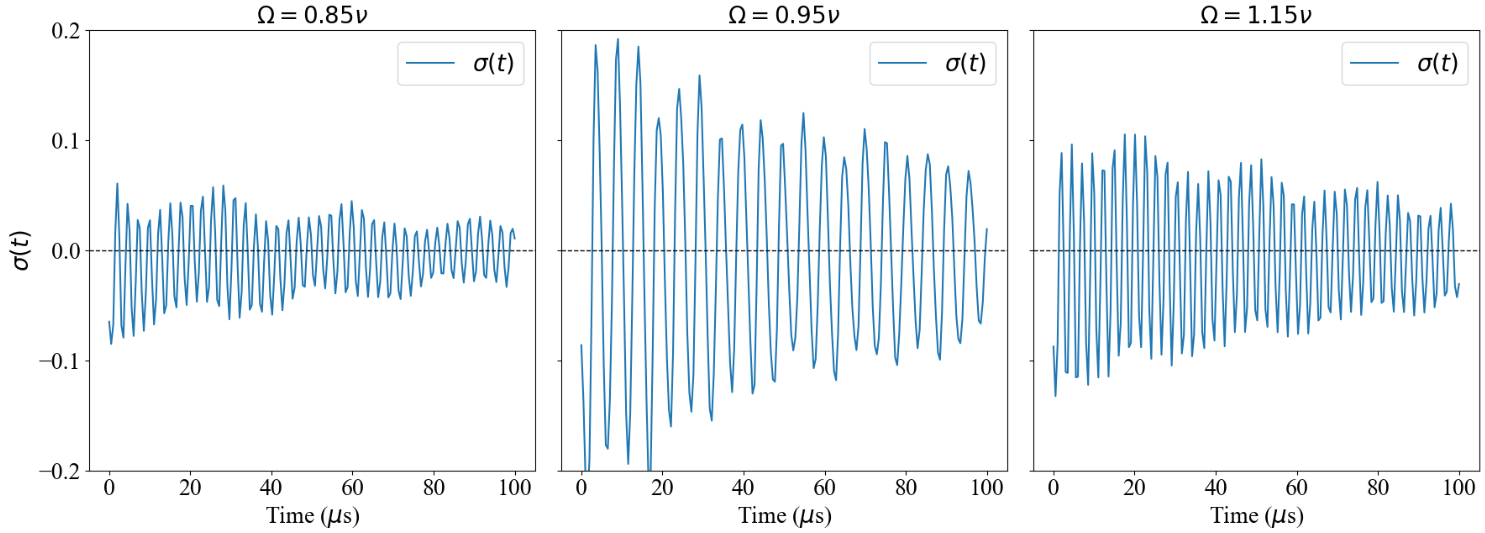}
\caption{Numerical results for $\sigma(t)$ under resonant conditions ($\delta = 0$). Positive values (above the dashed line) indicate non-Markovian dynamics. The system is initialized in $\rho_1 = \ket{g}\bra{g}$ and $\rho_2 = \ket{e}\bra{e}$. Other parameters are given in the text.}
\label{fig:S1}
\end{figure}

In Fig.~\ref{fig:S1}, we show the time evolution of $\sigma(t)$ defined in Eq.~(\ref{sigma}), for the initial states $\rho_1 = \ket{g}\bra{g}$ and $\rho_2 = \ket{e}\bra{e}$, corresponding to eigenstates of $\sigma_z$. The dynamics is computed over $100~\mu$s with a temporal resolution of $0.5~\mu$s. We consider resonant conditions ($\delta = 0$), with $\eta = 0.069$, $\gamma = 0.004$, and a motional mode of frequency $\nu = 2\pi \times 2.32~\text{MHz}$, initially prepared in a thermal state with $\langle n \rangle = 0.05$. The coupling strength is set to $\Omega = 0.85\nu$, $0.95\nu$, and $1.15\nu$. Positive values of $\sigma(t)$ signal non-Markovian dynamics. As shown in Fig.~\ref{fig:S1}, the positive oscillations of $\sigma(t)$ become more pronounced as $\Omega$ approaches $\nu$ (here with $\delta = 0$), indicating an enhancement of non-Markovianity near the resonance condition $\Omega \simeq \nu$. Away from this regime, for both $\Omega < \nu$ and $\Omega > \nu$, the amplitude and frequency of positive intervals decrease. This behavior is consistent with the results presented in the main text using eigenstates of $\sigma_x$.

\begin{figure}[t]
\centering
\includegraphics[width=\linewidth]{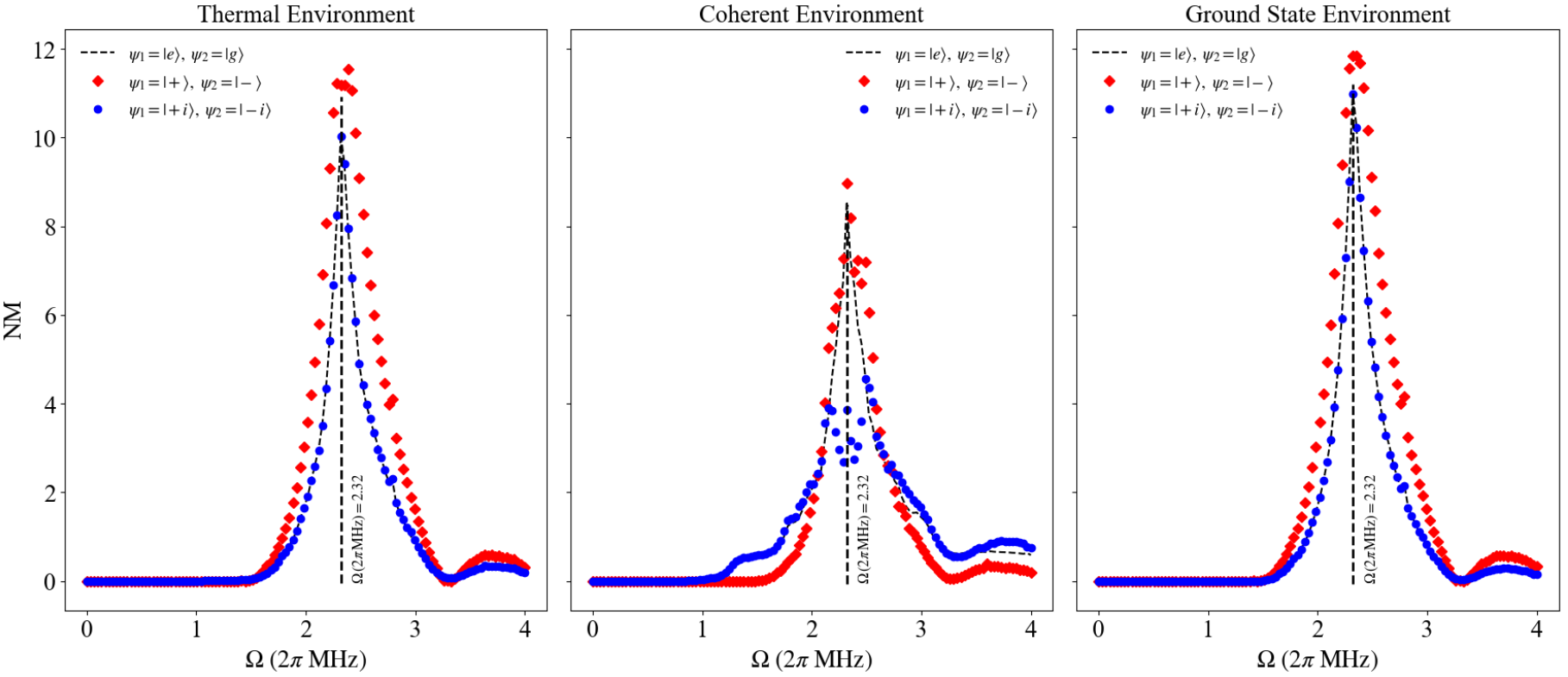}
\caption{Non-Markovianity measure (Eq.~(\ref{NM}), without maximization) evaluated for different initial spin and motional states, using the same parameters and time interval as in Fig.~\ref{fig:S1}. Results are shown for pairs of orthogonal spin states: eigenstates of $\sigma_z$ ($\ket{e}, \ket{g}$, dashed lines), $\sigma_x$ ($\ket{+}, \ket{-}$, diamonds), and $\sigma_y$ ($\ket{+i}, \ket{-i}$, circles). The three panels correspond to different initial motional states: thermal with $\langle n \rangle = 0.05$ (left), coherent with $\langle n \rangle = 1$ (center), and the motional ground state (right).}
\label{fig:S2}
\end{figure}

In Fig.~\ref{fig:S2}, we present the non-Markovianity measure defined in Eq.~(\ref{NM}), evaluated without the maximization over initial states. The calculation is performed using the same parameters and integration time window as in Fig.~\ref{fig:S1}. We compare the results obtained for different pairs of orthogonal initial states, corresponding to eigenstates of $\sigma_x$, $\sigma_y$, and $\sigma_z$, and for different initial motional states. We find that the non-Markovianity exhibits similar overall trends across all cases, with only minor quantitative variations. In particular, a nonmonotonic dependence on $\Omega$, with enhanced values in the vicinity of $\Omega \sim \nu$, is generally observed, especially for thermal states with low occupation, as in our experiment, and for the motional ground state. These results support the use of a fixed pair of $\sigma_x$ eigenstates in the main text as a representative and experimentally convenient choice.

\subsection{From Carrier-Dominated to Strong-Field Dynamics}
\begin{figure}[h]
\centering
\includegraphics[width=0.95\columnwidth]{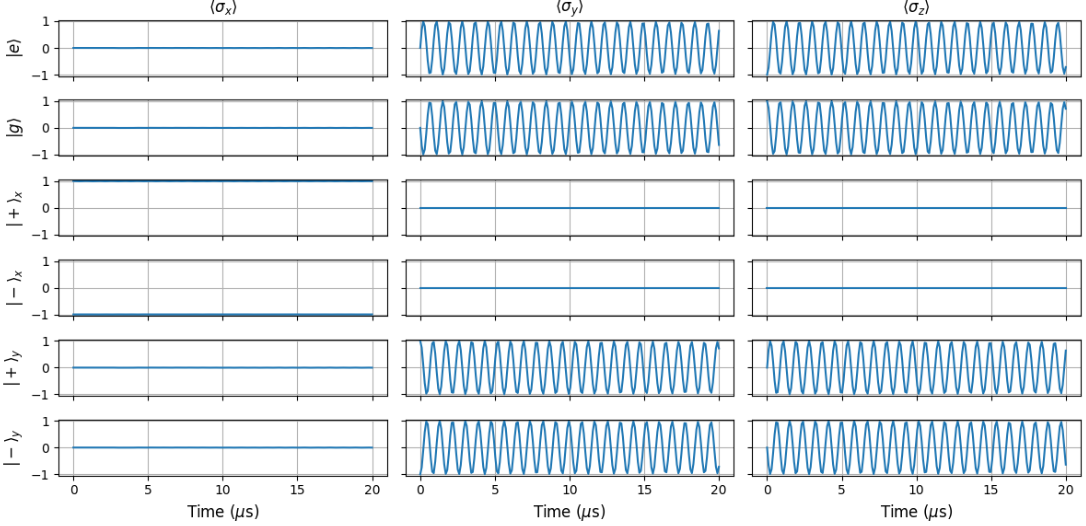}
\caption{Time evolution of $\langle \sigma_x \rangle$, $\langle \sigma_y \rangle$, and $\langle \sigma_z \rangle$ in the weak-field regime ($\Omega \ll \nu$), for different initial spin states: $\ket{e}$, $\ket{g}$ (eigenstates of $\sigma_z$), $\ket{+}_x$, $\ket{-}_x$ (eigenstates of $\sigma_x$), and $\ket{+}_y$, $\ket{-}_y$ (eigenstates of $\sigma_y$). Parameters are as in Fig.~\ref{fig:S1}, except for the mean motional occupation, here set to $\langle n \rangle = 0.09$, and $\Omega = 0.43\nu$. At $\delta = 0$, the dynamics is dominated by the carrier transition, resulting in an approximately constant $\langle \sigma_x \rangle$.}
\label{fig:weak}
\end{figure}

\begin{figure}[h]
\centering
\includegraphics[width=0.95\columnwidth]{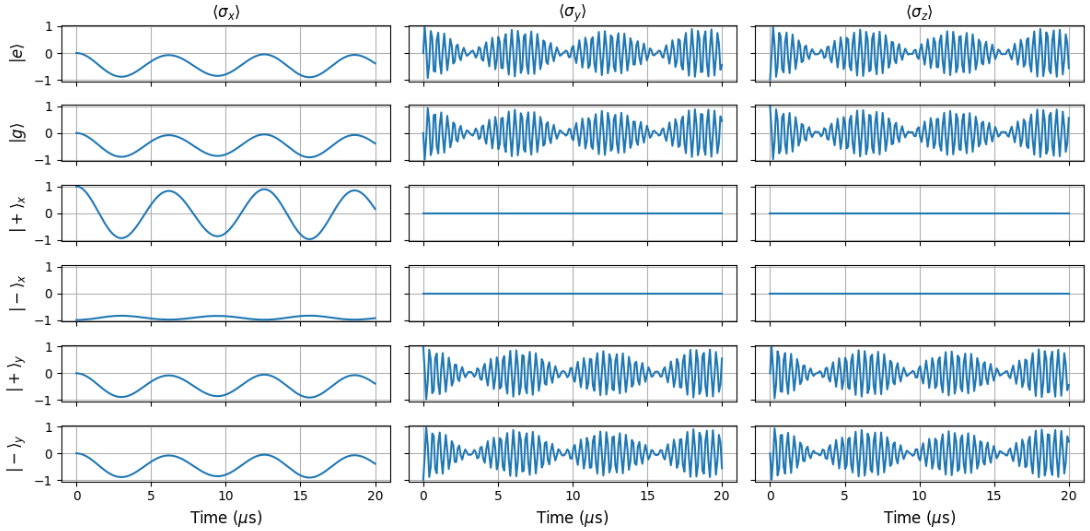}
\caption{Time evolution of $\langle \sigma_x \rangle$, $\langle \sigma_y \rangle$, and $\langle \sigma_z \rangle$ in the strong-field regime ($\Omega \sim \nu$), for the same initial spin states as in Fig.~\ref{fig:weak}. Parameters are as in Fig.~\ref{fig:S1}, except for the mean motional occupation $\langle n \rangle = 0.09$, with $\Omega = 0.95\nu$. In this regime, the dynamics departs significantly from the weak-field limit, exhibiting pronounced oscillations due to strong mixing between spin and motion.}
\label{fig:strong}
\end{figure}

To gain further insight into the pronounced differences in the dynamics under distinct driving regimes, characterized by the ratio $\nu/\Omega$, we analyze the time evolution of the qubit observables $\langle \sigma_x \rangle$, $\langle \sigma_y \rangle$, and $\langle \sigma_z \rangle$. We consider different initial qubit preparations, specifically the eigenstates of $\sigma_x$, $\sigma_y$, and $\sigma_z$, in order to compare how the driving regime affects the dynamics of these observables. We focus on the case $\delta = 0$. In the weak-field limit shown in Fig.~\ref{fig:weak}, one observes a nearly constant $\langle \sigma_x \rangle$. This behavior arises because, in this regime and for $\delta=0$, the dynamics is dominated by the carrier transition, described by an effective Hamiltonian proportional to $\sigma_x$ \cite{leib03}. In this limit, the evolution remains relatively regular and only weakly coupled to the motional degree of freedom. As the coupling strength increases (Fig.~\ref{fig:strong}), this simplified picture breaks down and the dynamics becomes significantly richer, with pronounced oscillations and strong mixing between spin observables due to enhanced spin–motion hybridization. These results highlight a qualitative change in the system dynamics as the coupling strength is increased from the weak-field regime ($\Omega \ll \nu$) to the strong-field regime ($\Omega \sim \nu$), which is experimentally accessed in this work.


\bibliography{Combined_PRR}

%
%
%
%
%
%

\end{document}